# Space telescope design to directly image the habitable zone of Alpha Centauri

Eduardo A. Bendek[a], Ruslan Belikov[a], Julien Lozi[b], Sandrine Thomas[d], Jared Males[d], Sasha Weston[a], Michael McElwain[e].

[a]NASA Ames Research Center, Moffett Field, CA 94035, USA
[b]National Astronomical Observatory of Japan - Subaru Telescope, Hilo, Hi, 808-934-5949
[c]LSST Corporation, AURA/LSST, 950N Cherry Av, Tucson, AZ 85719
[d]Steward Observatory, University of Arizona, Tucson, AZ 85719
[e]NASA Goddard Space Flight Center, Greenbelt, MD 20771

## ABSTRACT

The scientific interest in directly image and identifying Earth-like planets within the Habitable Zone (HZ) around nearby stars is driving the design of specialized direct imaging mission such as ACESAT, EXO-C, EXO-S and AFTA-C. The inner edge of Alpha Cen A&B Habitable Zone is found at exceptionally large angular separations of 0.7" and 0.4" respectively. This enables direct imaging of the system with a 0.3m class telescope. Contrast ratios in the order of $10^{10}$ are needed to image Earth-brightness planets. Low-resolution (5-band) spectra of all planets, will allow establishing the presence and amount of an atmosphere. This star system configuration is optimal for a specialized small, and stable space telescope, that can achieve high-contrast but has limited resolution. This paper describes an innovative instrument design and a mission concept based on a full Silicon Carbide off-axis telescope, which has a Phase Induce Amplitude Apodization coronagraph embedded in the telescope. This architecture maximizes stability and throughput. A Multi-Star Wave Front algorithm is implemented to drive a deformable mirror controlling simultaneously diffracted light from the on-axis and binary companion star. The instrument has a Focal Plane Occulter to reject starlight into a high-precision pointing control camera. Finally we utilize a Orbital Differential Imaging (ODI) post-processing method that takes advantage of a highly stable environment (Earth-trailing orbit) and a continuous sequence of images spanning 2 years, to reduce the final noise floor in post processing to ~2e-11 levels, enabling high confidence and at least 90% completeness detections of Earth-like planets.

## 1. INTRODUCTION

The question about the existence of planets beyond our solar systems, or exoplanets, has been in the mind of the human civilization for thousands of years. In fact, around 300BC, Epicurus wrote in a letter to Erodotus: "There are infinite worlds both like and unlike this world of ours". Scientists have been modeling planetary formation and evolution in detail for decades, however the first exoplanet was discovered only 20 years ago. In 1995 Mayor and Queloz [1] found Pegasi 51b by utilizing Radial Velocity indirect detection[1]. At the moment of writing, there are 5335 planet candidates and 1564 confirmed planets [2]. As a result, the exoplanet scientific community has compiled enough statistical information to predict the demographics of earth-like planets around different star types. Latest statistics suggest that there could be up to 55% probability that any given G, F or K type star will have at least one planet with 0.5 and 2 Earth radii within the Habitable Zone of the star.

This recent increase in earth-like occurrence rates has enabled us to consider and specialized mission to search for "the pale blue dot" on selected targets and not only survey oriented as all previous exoplanet missions. A target-oriented mission can be designed for this goal facilitating direct imaging of the exoplanets, which is high-impact scientific. Direct imaging presents two main challenges:

mission can be designed for this goal facilitating direct imaging of the exoplanets, which is high-impact scientific. Direct imaging presents two main challenges:

*Contrast:* An earth-like planet around a sun-like star is about $1 \times 10^{10}$ dimmer than the host star, as a result it is necessary to remove diffraction effects of the optical system and any kind of optical aberrations to avoid light contamination in the discovery zone.

*Angular separation:* The larger the distance from the earth to the host star, the smaller is the apparent angular separation between the planet and the star, therefore requiring a larger telescope to resolve it. For example, an earth analog around a typical nearby star located at 10pc will have an angular separation of 0.1", which requires at least a 1.5m telescope resolve the planet in visible light.

To image a planet it is also necessary to suppress the light from the start and the diffraction ring created by the telescope aperture, which is typically done using a coronagraph. NASA has studied coronagraphic direct imaging missions for at least a decade such as The Exoplanet Coronagraph (Exo-C) [4], the Exoplanet Starshade (Exo-S) [5], and the larger WFIRST/AFTA (Wide Field Infrared Survey Telescope / Astrophysics Focused Telescope Assets). These missions are scheduled to launch not before 2025 decade. This kind of missions are designed to be capable of imaging habitable planets around the nearest 5-20 stars, they cost $1B or more and they are not currently funded.

We propose a mission concept called ACESAT that is focused on taking the first image of and earth-like planet around the Alpha Centauri A & B (αCen A&B) binary star system. This system present a unique opportunity for direct imaging missions because it is the nearest star system to us, as shown in Figure 1; Specifically, αCen A&B habitable zones span 0.4-1.6" in stellocentric angle ~3x wider than around any other FGKM star. In theory this enables a visible light telescope as small as 25cm, equipped with a modern high performance coronagraph, to resolve the habitable zone at high contrast and directly image any potentially habitable planet that may exist in the system.

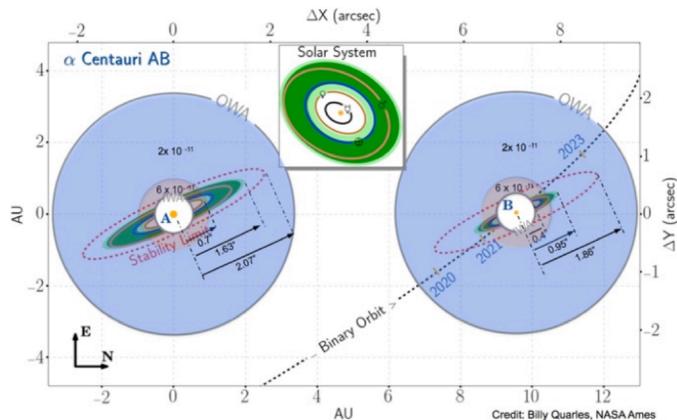

Figure 1: Apparent configuration of the αCen A&B system as seen from the earth. The disk on the left represents αCen A and the one on the right is αCen B. The HZ of each star is shown in green and light green to represent the classical and extended HZ.

According to the latest estimates from the Kepler mission, the occurrence rate of exo-Earths is about 55% per star (for size range of 0.5-2 Earth size and using the extended habitable zone definition [5]). Because αCen A&B is a binary, the chances of a potentially habitable planet existing around one of the two stars is 1- $(1-0.55)^2$ ~ 80% assuming independent probabilities. We realize that there is unique opportunity detect and observe for the first time an earth like planet in this system, and accomplish a high-priority directive of the decadal survey; *"The ultimate goal is to image rocky planets that lie in the habitable zone—at a distance from their central star where water can exist in liquid form—and to characterize their atmospheres."*

### 1.1 Scientific goals and objectives

The top-level scientific goal of ACESat is to directly image 0.5 to 2 earth radii planets' equivalent brightness, in the HZ of the αCen A&B planetary system during a two year mission. The specific science objectives are summarized below:

*Objective 1.1.* **Directly image planets** in the region around αCen A&B from the inner HZ edge out to the dynamical stability limit, with >90% completeness down to Earth brightness

*Objective 1.2:* **Directly image a debris disk** around αCen A and αCen B if one exists, (down to 1 zodi variations). A thick or more dense debris disk scatters starlight decreasing the plant contrast with background and requiring more signals for the detection.

**Objective 2.1. Determine the orbits of planets** found in obj. 1.1. by fitting a keplerian motion to the detected source over time.

**Objective 2.2. Constrain the size and mass of planets.** The dataset for this objective is the same as for objective 1.1, except that the reflectivity as a function of wavelength can be used to help constrain planet size.

**Objective 2.3. Characterize albedo variations**

The brightness of that planet will be measured on each image. A corresponding periodogram will also be computed that can suggest the planet rotation period.

## 2. MISSION REQUIREMENTS

To accomplish the scientific objectives we are interested in observing the HZ of both systems, from the inner edge found out to the stability limit where planets can remain in their orbits. These boundaries have been calculated at 0.93AU and 0.53AU for the inner HZ [5,6] and 2.79 +/- 0.65AU and 2.49 +/- 0.71AU for the stability limit [6] of αCen A and B respectively. Figure 1 shows the equivalent of these physical parameters in angular separation as seen from the earth, and for the most likely planetary system inclination. The classical and extended HZs are shown in dark and light green respectively. Also, a schematic of the solar system apparent size if placed at αCen A&B system distance is shown. Finally, this figure shows the areas accessible for the instrument baseline design in blue and red for different contrast levels. These values are summarized in table 1. The figure and the table refer to the Inner Working Angle (IWA), which is the smallest angular separation that the instrument can observe from the star, and the Outer Working Angle (OWA), which corresponds to the larger angular separation that instrument can observe.

Requirements shown in table 1 are in units of angular separation in $\lambda/D$, where $\lambda$ is the operational wavelength (500nm) and D, which is the telescope aperture (45cm) for this.

## 3. INSTRUMENT DESIGN

### 3.1 Instrument design rationale

The ACESat design is based on a compact space-based off-axis telescope with a high-performance loss-less internal coronagraph embedded as the secondary and tertiary mirrors. The light reaches the scientific camera after only 5 reflections maximizing throughput and optical stability. Ground based telescopes cannot deliver the contrast requirement due to atmospheric perturbations, and day/night observation cycling regardless of theirs aperture.

*Top-level requirements*

The ACESat instruments' top-level requirements flow from the scientific goals 1.1 to 2.3, and the geometry and characteristics of the αCen A&B systems. They can be summarized as achieving $10^{-8}$ raw contrast in the region of interest that spans from 1.6 to $12\lambda/D$ over 5 bands at 10% width covering from 400 to 700nm. This mission relies in the ability of enhancing the instrument raw contrast ratio of $10^{-8}$ by a factor of $5 \times 10^2$ raw to achieve a final contrast of $2 \times 10^{-11}$ utilizing calibration and data obtained during 30 days of continuous integration. This should allow to detect an earth-like planet with SNR=5.

The instrument contrast CBE is $0.5 \times 10^{-11}$, assuming an exposure time of 30 days per star per quarter, a raw contrast of $1 \times 10^{-8}$, a 45cm aperture telescope, and end-to-end efficiency that ranges from 46% to 58% (considering mirror reflectivity, detector QE, and dichroic losses) as a function of wavelength, and ODI data post processing. This contrast allows an SNR=20 detection of an earth like planet with ample margin with respect to the requirement of SNR=5.

### 3.2 Instrument overview

The ACESat instrument functional block diagram and optical design is shown in Figure 2, which is comprised of the Instrument Element and the Spacecraft Element (S/C). There are three instrument subsystems: a) *Shroud Assembly (SA)*, b) an *Optical Telescope Assembly (OTA), and* The *Focal Plane and Electronics Subsystem*. The OTA includes: a) a Telescope, b) a high-performance coronagraph, c) a Deformable Mirror (DM), d) a Focal Plane Occulter (FPO). The *Focal Plane and Electronics Subsystem* is composed of: a) Focal Plane Assembly that includes the Science detector and the Low Order Wave Front Sensor (LOWFS), both mounted to the OTA and b) the Payload Electronics Box (PEB) that contains the controllers for the DM, detectors, and tip/tilt, as well as the payload computer is located within the S/C. The *Optical Telescope*

Table 1: ACESat scientific requirements

| Contrast | IWA | OWA |
|---|---|---|
| **aCen B** | | |
| $6 \times 10^{-11}$ | 0.4" | 0.95" |
| $6 \times 10^{-11}$ | $1.6\lambda/D$ | $3.8\lambda/D$ |
| **aCen A** | | |
| $2 \times 10^{-11}$ | 0.7" | 1.63" |
| $2 \times 10^{-11}$ | $2.7\lambda/D$ | $6.5\lambda/D$ |
| **Stability limit (aCen A)** | | |
| $2 \times 10^{-11}$ | | 2.07" |
| $2 \times 10^{-11}$ | | $8.3\lambda/D$ |
| **Sensitivity** | | |
| SNR=5 | 1.6 Days | SNR=5 |
| ODI Calibration | 30 Days | ODI Calibration |

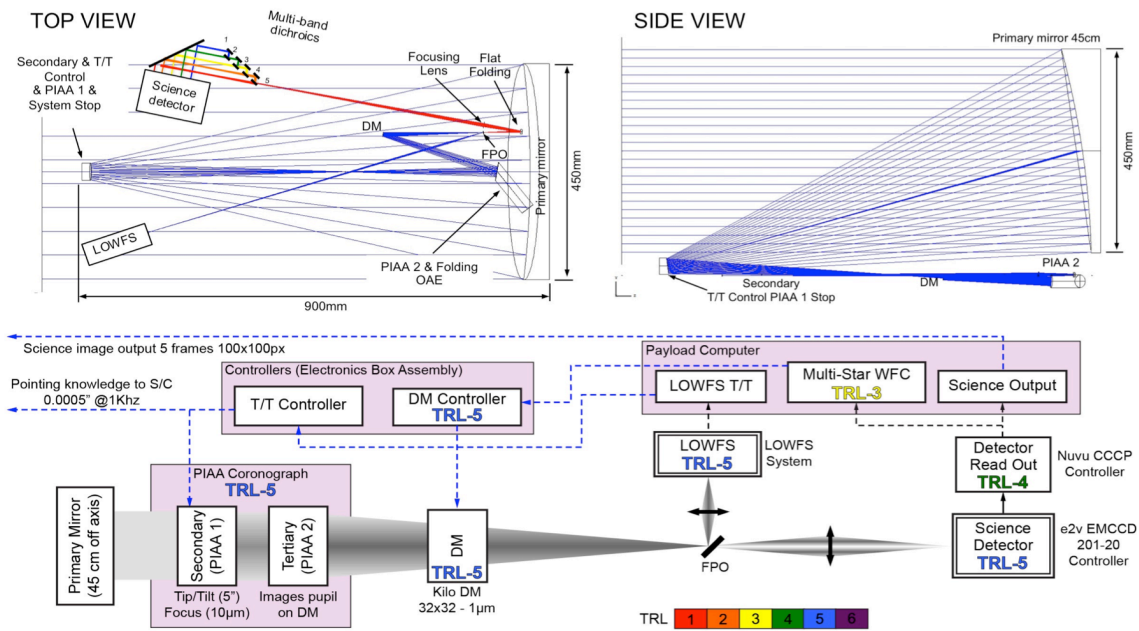

Figure 2: ACESAT subsystems and components and functions. Top: Instrument optical layout side and top view are shown. Bottom; Functional block diagram of the instrument including subsystem TRL assessment in subsystem areas

*Assembly (OTA)* utilizes a 45cm aperture off-axis Ritchey–Chrétien (RC) telescope that has a Phase Induced Amplitude Apodization (PIAA)[8,9,10] coronagraph embedded on the secondary and tertiary mirror. A real time computer reads the LOWFS data and sends control commands to the secondary mirror to control tip/tilt and defocus. The off-axis planet light continues through a series of dichroics creating 5 bands from 400 to 700nm, which are imaged by an EMCCD science detector. The mirrors are coated with protected silver for optimal reflectivity. The PEB, accommodated within the S/C Element, include the camera and actuator controllers, and the payload processors. A Shroud assembly provides stray light control, temperature control and optics protection functionality.

### 3.3 The Optical Tube Assembly

The OTA includes a Silicon Carbide telescope complaint with mechanical and thermal properties requirements, for mechanical hysteresis and distortions due to differential Coefficient of Thermal Expansion (CTE). The telescope is operated at 10˚C, and must remain stable for each observation period (1 quarter) to 1˚C along metering structure and 0.3˚C lateral gradients across the primary mirror, which is a 45cm aperture off-axis SiC 80% lightweighted. The secondary mirror of the OTA is also the first component of the coronagraph (PIAA 1), which acts as the system stop and provides tip-tilt/focus control. The secondary mirror active mount has a tip/tilt range of +/-5" and a focus range of 10μm. The tertiary mirror of the OTA is also the second component of the coronagraph (PIAA 2) and reimages the telescope pupil on the Kilo-DM deformable mirror, which is a 9.6 mm square with 32x32 actuators that have a 300 μm pitch and 1 μm stroke. The DM corrects the wave front errors creating a high-contrast dark zone where the planets can be discovered. Downstream of the DM the target star is imaged on the FPO surface, which acts as an angular beam splitter rejecting the star light into the LOWFS that provides tip/tilt and focus measurements. (Figure 3)

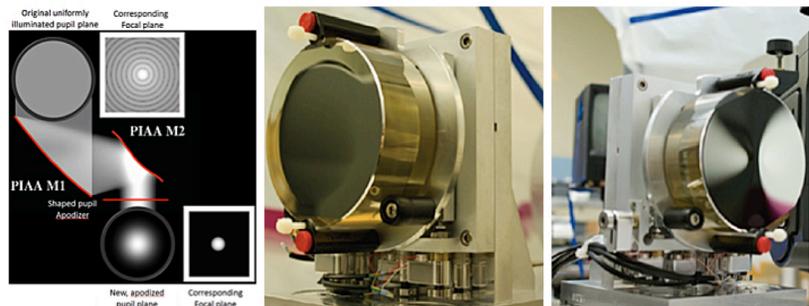

Figure 3. Representation of the PIAA coronagraph principle (left) and actual PIAA mirrors at NASA Ames Research Center (center, right).

### 3.4 Starlight Suppression System design

The coronagraph used by ACESAT is based on the Phase-Induced Amplitude Apodization (PIAA) coronagraph which members of the Ames Coronagraph Experiment (ACE) team have been pioneering[9,10] and maturing[11]. Without a coronagraph, the image of a star in a conventional telescope is the so-called Airy pattern, which is the Fourier transform of the telescope pupil. This Airy pattern has diffraction rings, which overlap the location of the habitable zone, and are many millions of times brighter than planets. PIAA eliminates these sharp edges by employing two aspheric mirrors. The design process of PIAA systems for unobscured apertures is very mature with multiple PIAA systems designed for several mission concepts apertures such as ACCESS, PECO, EXCEDE, and Exo-C, with the PI and other members of ACESat being directly involved with these designs. The ACESat PIAA design will be optimized during phase A trade study, but as a baseline and an existence proof, we adopted a design similar to the Exo-C PIAA, which in the absence of errors achieves a contrast better than $1\times10^{-8}$ averaged between 1.6 and $10\lambda/D$, satisfying ACESat requirements.

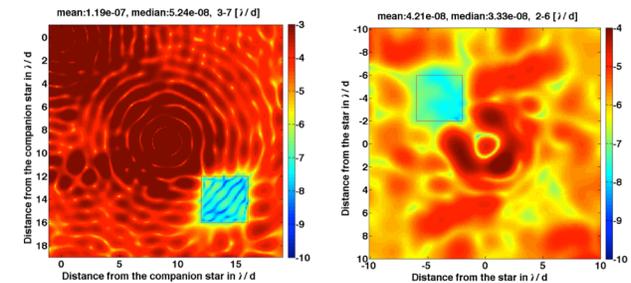

Figure 4. Left: Demonstration of MSWC on the left where a zone of high contrast can be achieved by controlling speckles from a binary component (top left) folded outside the nominal AO control radius. Right: Simulation of the algorithm applied the Alpha Cen

**Multi-star wavefront control and second star supression.** Currently Electric Field Conjugation (EFC) [11] or similar algorithms has been used at the ACE and JPL PIAA coronagraphic testbeds to measure the coronagraph wavefront and creating dark zones controlling the DM. For binary systems like Alpha Centauri, incoherent light coming from the (off-axis) companion diffracts light in the dark-zone of the target on-axis star. Moreover, the angular separation of the binary system will be beyond DM control region of 16 $\lambda/D$, limited by the number of DM actuators, for band 1 on the first year of the mission (31.2 $\lambda/D$) and all the bands by the end of the mission due to increasing angular separation of the binary system.

Implementing the Super Nyquist Multi Star Wavefront Control (MSWC) [12,13,14] solves the dark zone and angular separation problems. MSWC utilizes PSF replicas created by the DM quilting to control dark zones beyond the Nyquist frequency of the DM. This algorithm is currently about TRL-3 thanks of rapid progress achieved at the ACE Laboratory as part of an APRA effort funded to develop this technology. The demonstration of the Multi-Star algorithm is shown in Figure. The dark zone with a $5\times10^{-8}$ contrast is created at 3 $\lambda/D$ for the target star with a binary of the same intensity located at 14 $\lambda/D$.

The SNWC algorithm was also demonstrated for the ACESat science case for year 2022 observing on band 3 (lambda=555nm at 10% bandwidth) as shown on Figure 4. The separation between the 2 targets is 29 $\lambda/D$ and the potential planet at 4 $\lambda/D$ from the on-axis star. A dark zone of $3.3\times10^{-8}$ median contrasts is created by the DM between 3-7 $\lambda/D$ from the on-axis star. For these simulations we used a classical apodized coronagraph, ensuring that the results will be significantly better when a better performance PIAA coronagraph is modeled in the wavefront control simulation.

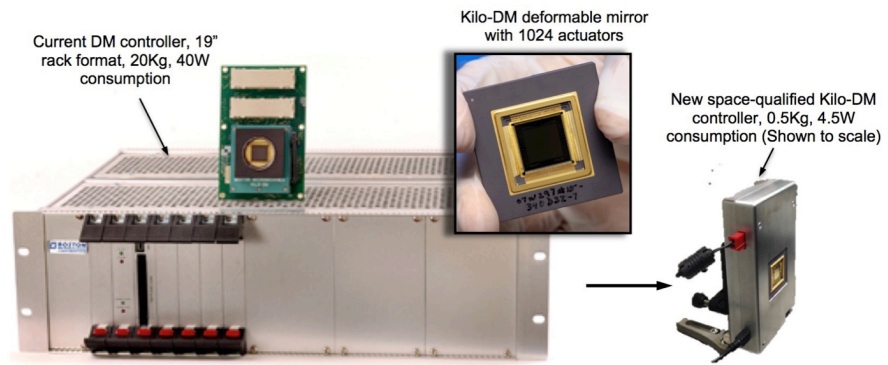

Figure 5. Old and new DM controllers. Flight prototype demo shown on the right.

**Deformable Mirror (DM).** A DM is used to remove speckles created by low special frequency figure errors in the mirror surfaces. We selected a Boston Micro Machines Kilo DM, which has 1024 actuators with a 300 μm pitch and 1 μm stroke for the space-qualified version. The DM has a square 10x10 mm aperture. Currently, the DM controller is large, heavy and not space qualified. We are developing an integrated wave front control system that contains the DM and the controller in a single solid-state package. This system weight is only 0.5 Kg and consumes 4.5 W. The demonstration controller shown on Figure 5 has been developed to demonstrate feasibility. The development cost has been included in the budget. Also the DM maintains a $\lambda/25$ @ 550 nm wave front error on the focal plane the telescope uses a DM to compensate for dynamic errors induced by thermal loads.

### 3.5 Instrument fine pointing control and jitter stabilization

The instrument has camera that receives the star light reject by the coronagraph and measures the payload jitter and stabilizes it within 0.5 mas by sending Tip/tilt and defocus commands to the actuator placed behind the secondary mirror. A Low Order Wavefront Sensor (LOWFS) utilizes a 3-zone focal plane mask that rejects the core of the star's PSF to an imaging system that magnifies the PSF for proper sampling on a fast camera that takes slightly defocused images of the PSF, running at 1,000 Frames Per Second (FPS). Our baselined camera is the Imperx Bobcat CLB-B0610M-TC, with a True sense CCD KAI-0340S from which we read out only 40x40 pixels. This configuration can deliver a pointing knowledge better than 1e-3 $\lambda/D$ RMS. The algorithm used to calculate the correction is an LQG [15] controller, which uses a Kalman filter. CBE stability is better than 1.5e-3 $\lambda/D$ rms, i.e. 0.00038"@550 nm.

### 3.6 Mechanical and thermal design

The OTA mirrors and metering structure will be manufactured of SiC given its low thermal distortion, high stiffness, high optical quality, and dimensional stability needed for ACESat. The SiC manufacturing process allows the manufacturing of extraordinarily complex parts that have virtually zero shrinkage (less than 0.1%) allowing to cast the part to its near-net shape reducing risk and saving cost and schedule.

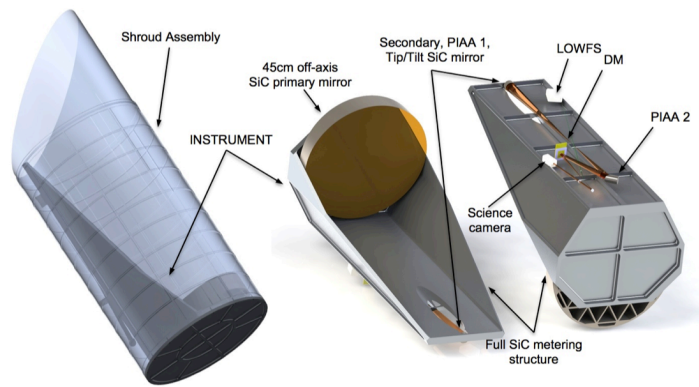

Figure 6: ACESAT subsystems and components and functions. Top: Instrument optical layout side and top view are shown. Bottom; Functional block diagram of the instrument including subsystem TRL assessment in subsystem areas

SiC optical substrates undergo a surface generation process using state-of-the-art 3-axis computer controlled grinding machines. The first few passes achieve the substrates mass margin and then the following passes create the specified surface figure prior to optical polishing.

The Silicon Carbide OTA is based on an L-Shaped structure that can support the Off-axis optics as well as provide a uniform surface to avoid thermal gradients (Fig. 6). The structure has be been optimized to minimize weight and maximize stiffness. The OTA weights 25.2kg including the metering structure and mirrors. Its first natural frequency is at 138Hz which is higher than any vibration source on the S/C avoiding coupling. The OTA will be attached to the S/C mounting plate utilizing 3 Semi-kinematic titanium flexure mounts and lapped G-10 pads to thermally isolate the OTA and limit mount distortion and bolted and pinned joints maintain alignment under launch conditions. A polished aluminum shroud that is covered with MLI for insulation is also attached to the S/C mounting baseplate.

The instrument will be maintained above its environment at 10˚C using proportionally controlled polyimide strip heaters. These heaters will be divided into six zones and be mounted to the top and back sides of the silicon carbide structure (Fig. 7). The instrument thermal design requires a peak operational power of 19 Watts and a Safe Mode power of approximately 4 Watts.

The CCD will be thermally isolated (with G10 spacers and a low emissivity finish) and thermally coupled to a dedicated CCD radiator on the sun shield capable of maintaining the CCD at -85 ˚C. Other heat dissipating components on the coronagraph will be thermally isolated and heat strapped to a heat pipe, which will be

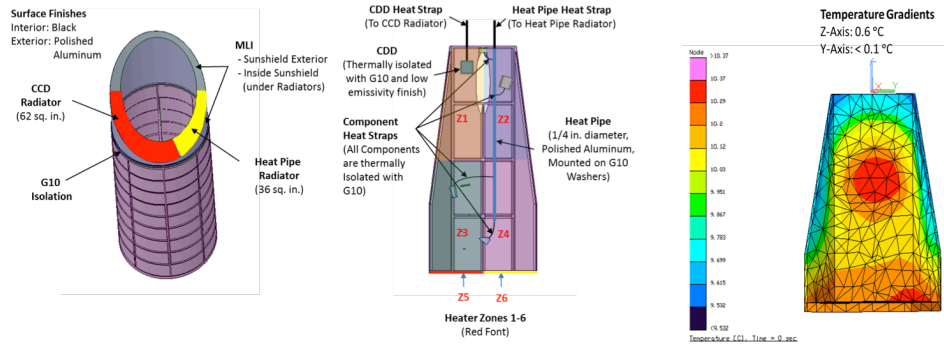

Figure 7. Instrument thermal control system and thermal analysis showing the coupled to a second radiator on the sun shield. The sun shield (extending above the spacecraft top deck) will be blanketed on all external surfaces including under the radiators. The sun shield shroud will have a polished aluminum exterior surface finish and will also be isolated from the aluminum cylinder with a G10 spacer. The instrument thermal design meets equilibrium (10˚C), axial ˚C and lateral temperature requirements with margin (Figure 6).

### 3.7 Focal Plane and Electronics

Focal plane detectors and their associated electronics accompany both the LOWFS and the science camera. These functions performed are photon detection, sensing, storage, and control of the WFS&C system. The payload computer executes the computational processes for these functions.

**Science Detector (SD) and Dichroic System:** The requirements to select the scientific detector are Quantum Efficiency (QE) better than 70% from 400 to 700nm, less than $0.05e^-$ Read Out Noise (RON) and a Dark Current (DC) less than $5 \times 10^{-2}$ $e^-$/pix/s to be able to reach the SNR or sensitivities required exposure times are 10s or shorter to prevent more than 1% of the frames from being affected by cosmic rays, imposing a very low read out noise requirement to avoid a significant penalty on RON.

We selected the e2v 201-20 Electron Multiplier CCD (EMCCD) which outperform the requirements achieving $0.03e^-$ RON and $5 \times 10^{-4}$ $e^-$/pix/s DC. The detector has a format of 1024x1024 pixels with 13μm pitch, which matches the desired sampling of 4pix per resolution element (resel) to perform EFC and MSWC. Our resel is 0.31" resulting in 51pix per HFoV or 100x100pix for each band. The five bands are filtered using sequential dichroics and mirrors as shown in Figure 8, which are imaged adjacent to each other on the detector to avoiding moving parts for the band selection. All the bands will be imaged simultaneously, however only one band will be actively controlled by the DM at a time. Nuvu manufactures cameras based on the EMCCD 201-20 detector [16] and provides a controller that achieve the desired performance (Fig. 8)

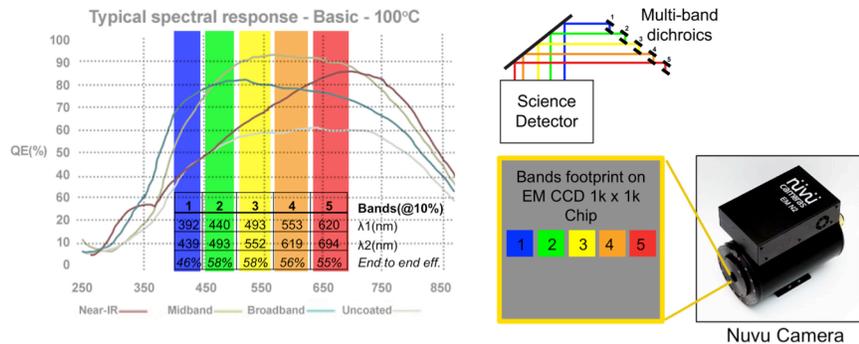

Figure 8. Left: QE of the Science detector e2v EMCCD 201-20 over-plotted by the five spectral bands selected. Top Right: a schematic of the multi-band dichroic system that operates in a converging beam and provides equal path lengths to the detector to ensure confocality between bands. Bottom Right: A depiction of the band locations on the detector and a Nüvü camera and controller that uses the EMCCD 201-20.

The same system has been selected for the AFTA-C mission and a technology development effort with Nuvu, the Center for Electronic Imaging, and First Light Imaging have commenced to bring these detectors to TRL-6 by January 2016. ACESat will follow this detector maturation program and implement the proven version on the mission. The CCD will be cooled down to -85˚C utilizing thermoelectric cooling and a conventional radiator.

**LOWFS Camera**: The Low Order Wavefront Sensor (LOWFS) measures the payload jitter and stabilizes it within 0.5 milliarcseconds by sending tip/tilt and defocus commands to the actuator placed behind the secondary mirror. The LOWFS utilizes a 3-zone focal plane mask that rejects the core of the star's PSF to an imaging system that magnifies the PSF for proper sampling on a fast camera that takes slightly defocused images of the PSF, running at 1000 frames per second (FPS). An off the shelf camera with these capabilities is the Imperx Bobcat CLB-B0610M-TC with a True sense CCD KAI-0340S that has 7.4um pixels. The LOWFS reads out a 40x40 pixel region of interest on the detector. This configuration can deliver a pointing knowledge better than 1e-3 $\lambda/D$ RMS. The algorithm used to calculate the correction is an LQG [17] controller, which uses a Kalman filter and a priori knowledge of the disturbance, obtained by a procedure of identification of the vibrations. With this design, a stability better than 1.5e-3 $\lambda/D$ rms, corresponding to 0.38 mas at 550 nm, can be achieved.

### 3.8 Instrument Software and Electronics

The instrument is operated by a payload computer that perform the following tasks; Execute the payload master control routine that computes the wavefront control and Jitter control commands, as well as pre-process and deliver science products. This computer also run the DM, LOWFS, Science camera, heaters controllers and the communications with the S/C.

Our data rate output is defined by the following variables. 100x100px frames sampled at 16bit resulting on 20KB frames. We sample 5 bands simultaneously generating 100KB per exposure. The exposure time has been set to 10s to avoid that a cosmic ray damage (3-4 hits/cm^2-sec) a long integration frame. An onboard algorithm using a median filter will remove frames corrupted by cosmic rays or high energy particles before storing them.

A control electronic box will be mounted separately from the instrument to avoid heat dissipation instabilities. This box will contain the payload computer and controllers for the two cameras, DM, heaters and Tip/Tilt secondary. We estimate a total power consumption of 40W on this box, that will be mostly dissipated on the box thru a thermal connection to the S/C.

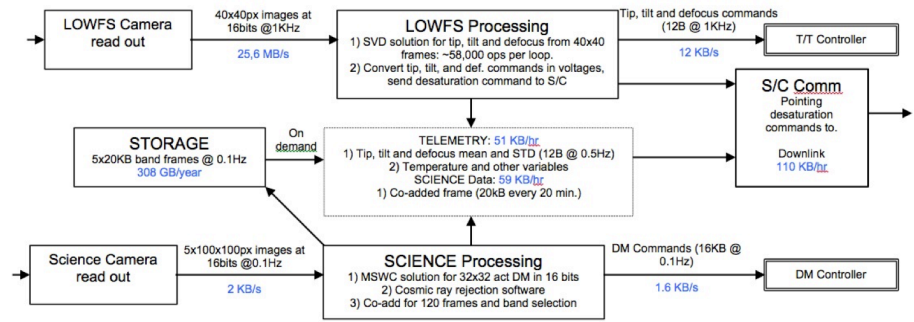

Figure 9. ACESat software functional diagram

The payload computer processing power is dominated by the LOWFS sensor that needs to perform 1,000 Singular Value Decomposition of 40x40px frames (M=1600px) into 3 modes (N) per second, resulting in 5.8x10$^7$ float point operations per second (Based on a SVD complexity of $4MN^2+8N^3$). In addition, Multi-Star Wave front Control has similar processing powers requirements and it will be executed every 10 seconds. This algorithm requires 64GB of memory to store large matrices needed to run the EFC routine. The Software functional diagram is shown in Figure 9.

The Payload Instrument suite utilizes three boards from MOOG BROAD REACH for managing controllers, facilitating the closed loop LOWFS control system, and Science Detector data management: The AJEET BRE 440, the CMOAB (4 GB RAM), and the DMOAB (12 GB Flash).

### 3.9 Radiation, contamination and performance degradation

The nominal radiation environment for the ETO in this period is predicted to be 6krads for the baseline mission duration. S/C and Instrument electronics will be assessed to carry significant margins beyond this to ensure operation for the full mission. The design performances consider end-of-life specifications.

A contamination control plan will be implemented for material selection (Adhesives, coating, etc.), design (eg. Shroud forms a CC barrier, and for I&T to ensure particulate and molecular contamination are controlled to acceptable levels, nominally by conducting I&T activities within a 10k cleanroom environment, with appropriate controls (personal & procedures) and protocols (eg. Bake-outs). A plume analysis will be performed to assess On Orbit contamination during operations. Degradation affects will be consider when selecting EEE parts such as CCD's and FPGA's.

Table 2: ACESAT interface values

| Resource | CBE | [units] |
|---|---|---|
| Instrument mass | 45.4 | kg |
| Power | 60.5 | W |
| Voltage (DC) | 100, 28 &, 5 | V |
| SC Pointing (1-Sigma) | 5 | arcsec |
| Sci. Ops. data rate | 120 | KB/hr |

### 3.10 Instrument MEL and Budgets and ICD

The instrument has a total mass of 45.4kg. The SiC OTA weights 25.2kg and remaining is electronics, harness and cables. The PEB consumes 40.5W and 20W addition power is allocated for heaters with a total of 60.5W. Table 2 shows the interface values for mass, power, pointing and data rates required.

## 4. SPACECRAFT DESIGN

The spacecraft is a high-heritage, low risk design, It incorporates only subsystems and components used by NASA-Ames or SSL on recent spaceflight missions. The S/C is designed to provide sufficient Delta-V for injection into an ETO orbit, provide payload accommodations including Power, TTC, TCS and pointing stability to meet mission objectives.

### 4.1 Subsystem descriptions

**Structure**: The spacecraft primary structure design is driven by a desire to have the first structural vibrational mode at as high a frequency as practicable, to maximize separation from the frequency band of the pointing system. To accomplish this, we have selected a structural design technology consisting of Graphite-Epoxy honeycomb panels bonded to Aluminum frames, fastened at the edges. The spacecraft general layout will follow the octagonal geometry and dimensions of the LADEE design, extended in length to accommodate the telescope inside an enclosed volume. Solar array cells will be attached rigidly to the structural panels to avoid introducing any additional flexible modes. Internally, a modular design approach will be followed, with the 3 main modules being the propulsion module, the avionics module, and the payload module. A model of the structure is shown in Figure 10. The bus and the propulsion system is on the bottom of the S/C and the payload on top.

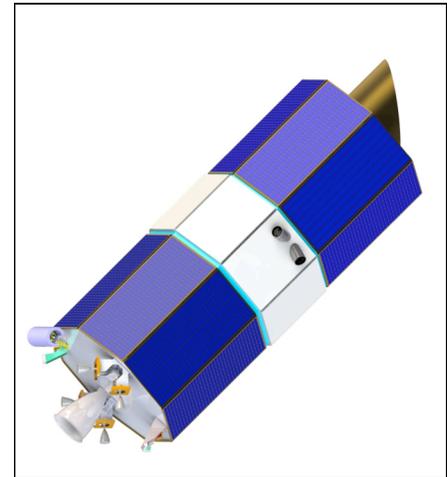

Figure 10. ACESat spacecraft and payload

**Propulsion Subsystem:** The Delta-v budget from GTO to Earth Trailing heliocentric orbit (C3 = 0.7) is 800 m/s; this requires a bi-prop propulsion subsystem. We have selected a build-to-print copy of the LADEE propulsion subsystem to provide this capability. It is a Helium pressurized, pressure regulated bi-prop system. It utilizes a single main thruster supplying 450 N of thrust with an $I_{sp}$ of 319 sec. The ACESAT Delta-v budget requires 93.4 kg of propellant (including 4.5 kg for ACS) which is stored in 4, 31 liter tanks including anti-slosh diaphragms. The system is sized to accommodate 134.2 kg of propellant (30% margin).

### 4.2 Command and Data Handling:

C&DH will be accommodated by a Broad Reach Engineering (BRE) Integrated Avionics Unit (IAU) of the same general characteristics as the IAU flown on LADEE, with modifications noted below. The LADEE IAU has a 3U form factor with a backplane accommodating up to 8 cards. The LADEE IAU populated the backplane with 7 cards, including a Single Board Computer (SBC) card hosting a Rad 750 CPU, a digital MOAB (DMOAB) card, an analog MOAB card (AMOAB), a Solar Array & Charge Control Interface card, 2 PAPI boards and 2 SATORI boards, leaving one spare slot. DC-to-DC converters are hosted on a separate panel that is heat-sunk to the IAU enclosure. The ACESAT IAU would be modified to include a Mass Memory card occupying the unused cPCI slot and providing 192 GB of memory. The DMOAB and AMOAB cards would each be modified to include an additional 12 GB of memory, bringing the total up to 216 GB

**Power:** The spacecraft nominal bus voltage is 28 VDC (21.5 to 32 VDC). Electrical power will be supplied by rigid, body mounted solar arrays attached to all sides of the spacecraft with 28.3% efficient solar cells will be used. The total array size is 7.2 m$^2$, which is sized to supply the steady state power requirement of 438.4 W (during all operational modes) at the end of life. The spacecraft battery is sized to achieve ≤ 80% depth of discharge during the worst case eclipse period (2 hours of 250 W power consumption during GTO burn). It is a build-to-print copy of the LADEE battery and consists of one, ABSL 24AH unit, incorporating LiFePo 18650 cells. The payload requires 100 VDC at low power to operate the steerable secondary mirror; this power will be supplied from a dedicated power board located in the payload integrated avionics box.

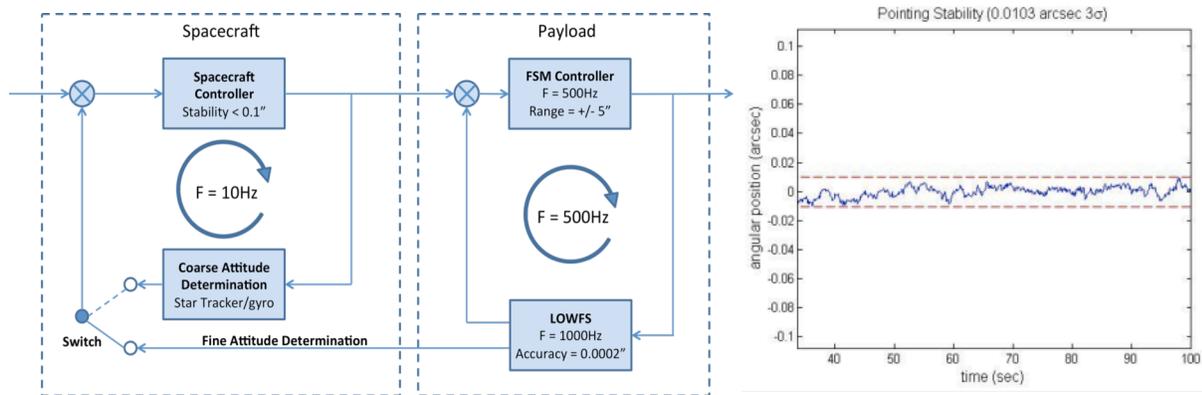

Figure 11. (Left) the strategy for the control system that will incorporate knowledge from the telescope and effect control with the RWA on the S/C bus; and (right) the stability achieved from a control system utilizing position and derived rate knowledge from the payload under the influence of constant solar torque.

### 4.3 Attitude Determination and Control (ADC):

*Spacecraft stability requirements*

The pointing stability requirement, which is driven by the PIAA coronagraph design, is 0.5 mas for all frequencies. The S/C pointing accuracy required is +/- 2.5" of the target in order for the instrument to acquire it. The payload LOWFS will provide common path, absolute pointing knowledge of 7.5x10$^{-4}$ arcseconds to the spacecraft attitude control and to the payload computer that will send correction tip/tilt and defocus corrections to the secondary mirror to maintain the target pointing within 1x10$^{-3}$ arcseconds. The ability of the Fine Pointing System of the instrument to control vibrations imposes the maximum jitter allowed on the spacecraft of 0.1" from 0 to 0.1Hz, an attenuation function with a slope of -3as$^2$/Hz. For 0.1Hz to 300hz, and 1mas jitter for frequencies higher that 300Hz.

**Table 3. Spacecraft pointing requirements**

| Pointing Accuracy | +/-2.5" |
|---|---|
| Jitter (RMS 1-σ between 0-0.1Hz) | 0.1" |
| Jitter (RMS 1-σ 0.1 to 300Hz) | -3as$^2$/Hz |
| Jitter (RMS 1-σ for >300Hz) | 0.5mas |

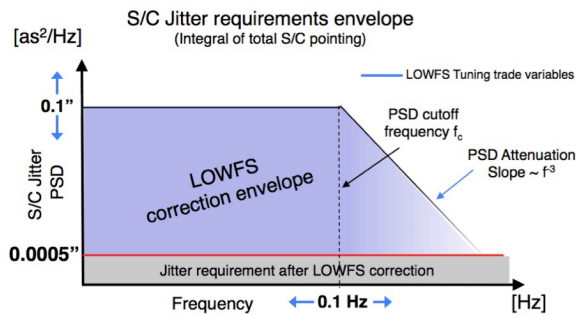

After an initial checkout and commissioning phase, the telescope will be pointed at the target star system and remain undisturbed for periods up to 100 days. This requirement drives the strategy of the ADC system design to a very low noise, disturbance-rejection type control system. The primary attitude pointing control actuator is a four-wheel tetrahedral-array Reaction Wheel Assembly (RWA). We propose to use 23 Nm-s Teldix wheels with a maximum torque of 0.09 Nm, each (SSL heritage). The RWA will be mounted to the spacecraft via a MOOG viscoelastic Vibration Source Isolation system to reduce wheel imbalance jitter transmitted into the spacecraft structure. The RWA wheel size is chosen to allow the storage of 100 days of solar torque without the need for desaturation. Desaturation of the RWA will be accomplished at the time of the quarterly spacecraft roll maneuver and will be effected by the bi-prop attitude control thrusters.

Initial simulation shows that deriving body rates from the telescope attitude at 10 Hz is acceptable for meeting the stability requirements levied on the spacecraft. Figure 11 right shows the stability achieved from a control system utilizing position and derived rate knowledge from the payload under the influence of constant solar torque.

The control system's stability is nearly a factor of ten more stable than the requirement. It should be noted that reduced jitter from the RWA mounted on dampers is not included in this model; however, SSL analysis for a similar wheel, without dampers, shows that this jitter will not exceed the stability requirement of 0.1".

**Thermal Control:** The spacecraft bus is a hot-side/cold-side, cold-biased passive design with strip heaters located near critical components. Spacecraft internal components will be maintained at 10C, +/- 10C. The spacecraft payload bay has a special controlled interface with the telescope. The spacecraft bus is required to provide 50 W of conductive thermal dissipation at the telescope mechanical interface and to maintain a radiative balance with the external surface of telescope shroud. Radiative balance will be maintained passively by keeping payload bay walls at constant temperature via albedo-controlling surface coatings and conductive heat straps. The spacecraft provides ICD-controlled data, power, thermal conductive, and thermal radiative interfaces to the payload which are critical to temperature management of the telescope metering structure.

### 4.4 Telecommunications Subsystem:

ACESat TT&C is driven by the concept of operations, which requires quarterly Ka and X band downloads of the science data at high data rates and weekly contacts with the spacecraft housekeeping (H&S) data downlink & Commands uplink. TT&C also operates during LEOP and spacecraft saving events. Four LGA's with 3db cone angles of 75° allow the spacecraft to contact Earth without reorienting the spacecraft, utilizing one LGA at a time via a switching network. Each LGA is used for both transmit and receive via a diplexer. The X-band transponder is capable of receiving command, transmitting telemetry and performing coherent ranging which would be needed for deep space missions. The telemetry output of the X-band transponder is connected to the X-band TWTA. The X-band downlink uses a 50-W TWTA. A link analysis, with 3db margin closes the link with 95% weather availability at 10 degrees of elevation over any DSN 34-m BWG antenna at the minimum data rate required for regular housekeeping telemetry. The quarterly science data Ka-band transmitter, horn antenna, and 35-W Ka-band TWTA provide an adequate EIRP. The Ka link closes with 95% availability at 20 degrees of elevation over any DSN 34-m BWG antenna for the minimum data rates required for quarterly downloads shown in Fig. F.2.4. The command signaling is in accordance with DSN standards for commanding. The downlink on the X-band will use BPSK modulation with subcarriers and NASA standard concatenated coding. The Ka-band downlink will employ BPSK modulation with (8920,1/6) turbo coding.

## 5. MISSION DESIGN

The baseline mission concept is for launch service delivery to a Geosynchronous Transfer Orbit (GTO). The Delta-V budget to transfer from GTO to Earth-trailing orbit is ≈ 800 m/s. This drives the spacecraft to specify a high performance bi-propellant propulsion system.

A survey of industry (via Sources Sought) indicated that no current or recent heritage spacecraft bus was available that would meet requirements 1-3 at a cost consistent with a Class D mission. A semi-custom design is therefore required, with heritage at the subsystem level. The proposed design is based heavily on use of LADEE and SSL communication satellite heritage spacecraft subsystems, including propulsion, avionics, structure, ACS, power, comm., thermal, harness, flight and ground software

ACESat will be powered-off during launch. ACESat will be commanded up to and including separation through the primary spacecraft. Separation sensing and confirmation will by ACESat.

Real-time command and telemetry communications will be performed using X-band. Ka-band will be used for science data. The NASA Near Earth Network (NEN) will be used while in GTO. During Heliocentric Orbit Insertion, the NEN and the Deep Space Network (DSN) will both be utilized. The DSN will be used in the operational orbit.

After a commissioning and calibration period, ACESAT will perform its primary mission observation over a period of two years. The telescope will be body pointed by the spacecraft. Coarse alignment will be through the spacecraft star-tracker.

ACESat will be placed into a helio-centric earth-trailing orbit for science observations. The orbital period is 367 days. In this orbit, the spacecraft will drift away from the earth at a rate of 0.21 AU per year. This orbit was selected since it offers a relatively benign environment (compared to earth orbit), minimizing spacecraft exposure to disturbances, and allows for continuous observation of target stars. The orbit parameters are shown in Figure 12.

### 5.1 Launch Services and Launch Vehicle Compatibility

Alternative access to space will be as a secondary payload with a commercial geosynchronous communications satellite as shown in Figure 12. The market for geosynchronous satellite is stable with a projected average of 23 GTO launches per year through 2020. The advantages of this secondary payload approach are similar to those for hosted payloads and include economic and schedule benefits.

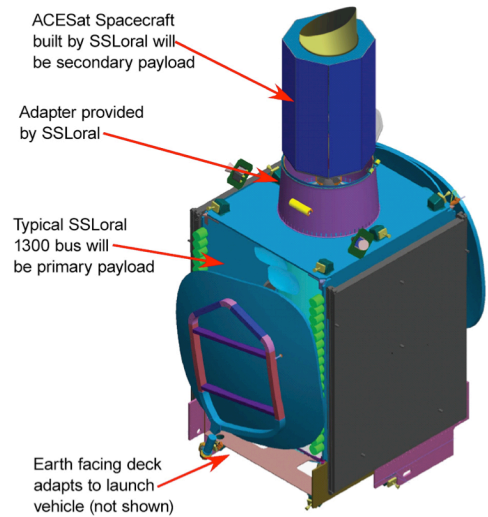

Figure 12: Rideshare arrangement on primary communications satellite

### 5.2 Operational modes and timeline

**Mission Timeline:** The nominal ACESat mission is divided into three phases–checkout, primary science, and extended science. The Checkout phase lasts approximately 90 days, and includes launch into GTO, heliocentric orbit injection, spacecraft checkout, instrument checkout, and dust cover ejection. The Primary Science phase lasts for the next two years and covers the observations of Alpha Centauri. The Extended Science phase lasts for one year and includes observations of Sirius, Procyon, and Altair. The instrument has the following operational modes that are the following:

a) *Off* – instrument is powered off. Survival heaters are on, controlled by the spacecraft
b) *Standby* - the instrument is powered, the payload processor is running and the subsystems are ready to perform operations. The instrument remains in this state during the quarterly downlinks and prior to the start of calibration.
c) *Calibration* – used during the calibration period.
d) *Target acquisition* – used to establish pointing following spacecraft coarse alignment on a target. (this mode is used at the start of each star observation)
e) *Data acquisition* - instrument is acquiring science data.

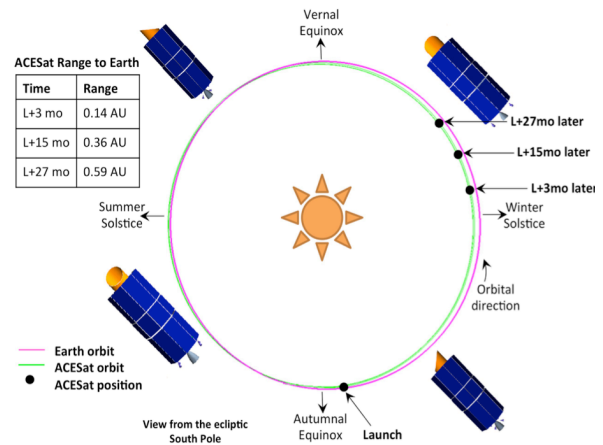

Figure 13: The ACESat S/C in each of its four orientations throughout the observing year.

**Target acquisition:**
*Step 1, Spacecraft slewing:* After a target is selected for observation, the S/C will slew to the target's coordinates and will hold the pointing to an accuracy of +/-2.5".
*Step 2, Spacecraft Jitter stabilization:* Then, the target will be within the LOWFS FoV, which will provide the instrument and the S/C with high-precision pointing information and the spacecraft will start a "low jitter" operation state.
*Step 3, Data acquisition.* The LOWFS commands the telescope secondary mirror stabilizes the pointing to the coronagraph down to 0.0005" during observations. Pointing update commands are sent to the S/C when drifts bring the tip/tilt range approaching the limit. MSWC will continue running at a gain low enough that the photon noise contribution of probes is negligible while still high enough to track slow variations of the quasi-static speckles, enabling 1e-8 raw speckle contrast over each quarter.

### 5.3 Calibrations

Once the telescope is on the desired orbit and pointed to the target we will calibrate the star brightness and adjust LOWFS exposure time, star alignment of the optical system and calibrate the LOWFS modes (tip/tilt and defocus). Then, LOWFS control loop will close to maintain the star within 0.0005" of the telescope optical axis. The next step is to select an observation band on the detector and execute MSWFC to achieve $1\times10^{-8}$ contrast in between 1.6 and 10 $\lambda/D$. This DM setting is saved for the next time that the same band is selected. After the calibration is performed the scientific target observations starts obtaining 10 second exposures for 1.6 days in each band.

*Science Observations/"Quarter-in-the-Life"*

Primary Science period will be divided into observing quarters, which last for approximately 91 days. During each quarter, the spacecraft will be inertially pointed at the target star, and will continuously collect data, until physical spacecraft constraints (sun avoidance, telecom, power, and thermal) dictate that the spacecraft to be re-oriented by rolling 90-degrees. Figure 13 shows the spacecraft in each of its four orientations throughout the observing year. There are no time critical events required to ensure the successful collection of science data. The ACESat operational cadence repeats on this quarterly basis, so it's useful to look at a "Quarter-in-the-Life" of ACESat.

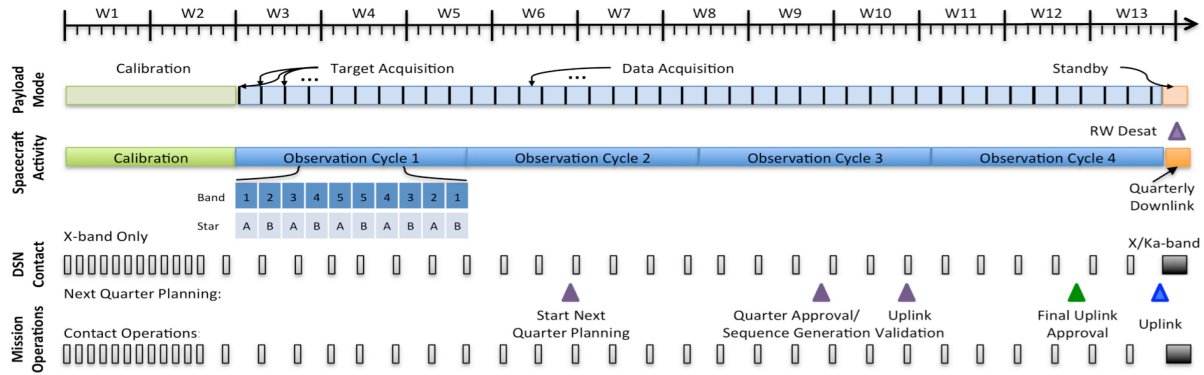

Figure 14: ACESat "Quarter-in-the-Life"

The science observation quarter is divided into three periods: Calibration, Science Observation, and Data Downlink. The quarter begins with the calibration period, which lasts for 14 days. During this period the pointing is calibrated, and final observing parameters for each observational band are determined. Science Observations last 75 days, and involve a series of observations of both stars, in all five bands of the camera. One complete observation cycle alternates viewing of the two target stars while stepping through each of the bands incrementally twice. This ensures for each cycle, each target star is observed once in each band. Figure 14. shows this "Quarter in the Life".

There are four observation cycles per quarter. Periodic Health and Safety (H/S) contacts lasting 6-8 hours are conducted every 2-4 days, and they utilize the X-band command/telemetry link via the low-gain antenna. During these contacts, real-time telemetry is acquired, and samples of the primary science images are downloaded to evaluate the health of the instrument. Other spacecraft housekeeping activities may be performed at the discretion of the Operations Team. Science observations continue through the H/S contacts.

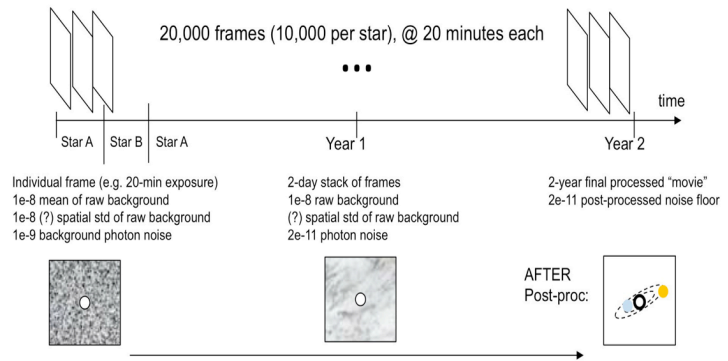

Figure 15: Schematic of the data and post-processing.

At the conclusion of the Science Observation period, there is a two-day Science Data Downlink period. Figure 16 shows a timeline of Downlink period's activities. During the Downlink period, the spacecraft is maneuvered to point the high gain antenna towards earth, to allow high data rate Ka-band transmissions to occur.

The following data is downlinked: Primary Science images, from the primary band; Stored Engineering data; Secondary Science images, from the other four bands; Retransmission of the data that not received on the ground to ensure completeness of the science data. Following the Science Data downlink, any required engineering maintenance activities are performed, including the desaturation of the reaction wheels. Next, preparations for the upcoming quarter are made, including uploading the command sequences for the quarter, and rolling the spacecraft. The Downlink period concludes with time for the spacecraft to achieve thermal equilibrium in its new attitude.

### 5.1 Data Sufficiency

The raw data from the mission will consist of an almost continuous sequence of 20-minute frames (co-added before download) spanning over 2 years, or 10,000 frames per year per star (2,000 per band per year per star) (see Figure 15).

Each individual frame will have a raw contrast of no more than our raw contrast requirement of 1e-8 from stellar speckle residual after MSWC (see Fig. 16). The speckle photon noise per 20-minute frame is 1e-9, which averages out to 2e-11 in a 2-day average, along with any other random noise at 1e-9 or lower contrast, uncorrelated between frames. Although these 2-day stacks will have photon noise well below Earth-like planet contrast, they will still have a bias (which will be different in each 2-day stack) that is up to 100 times brighter than the planet itself. Post-processing is necessary to remove this noise.

### 5.5 Post-Processing

There are three main types of residual noise on the image sequence: known speckle field coherent with the star; known speckle field incoherent with the star; and random variations. The first of these are systematic and the last one is random. We now cover each one in turn. ACESat post-processing makes use of a combination of techniques and is described in detail in [17]. The main principle that differentiates ACESat post-processing from other missions is that it has the benefit of tens of thousands of images spanning 2 years around the same star system. A special algorithm was developed called "Orbital Difference Imaging", or ODI [17] which leverage this unique feature of ACESat and enables a much greater post-processing factor than is available on missions that only have one or a few visits on a given target. A complete ODI reduction combines optimized PSF subtraction KLIP [18] and temporal low-pass filtering, along with spatial filtering and other standard image processing tools.

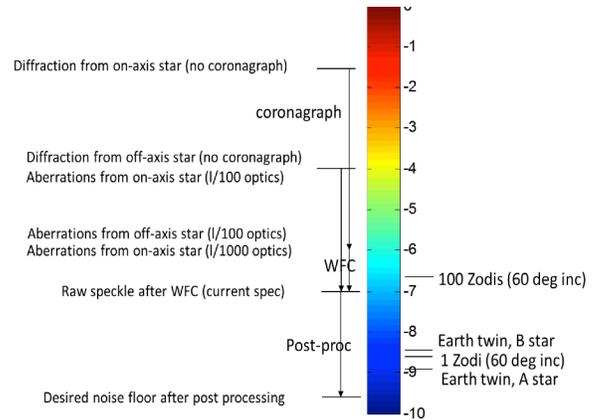

Figure 16: The contrast progression from acquisition to wavefront correction and post processing.

## 6. CONCLUSIONS

We have designed and instrument, spacecraft and mission to be able to image an earth-like planet in the HZ of the system αCen A&B. Based on our current performance CBE the mission will be capable of detecting planets equal or larger than $0.5R_\oplus$ assuming a typical albedo of 0.3. The discovery zone includes the inner HZ of both star that corresponds to 0.4" for αCen B out to the stability of both stars that corresponds to 2.07" for αCen A. The mission will be capable of achieving its science goal with 90% confidence and it is equipped with expendables to continue an extended science mission. The mission would fit on the cost cap and maturity of a NASA SMEX class and it should last two years with a possible extension of 1 more year to achieve the enhance science cases.

## ACKNOWLEGMENTS

This mission concepts is the result of the effort of large group of people from the NASA Ames Mission Design Center and industrial partners Loral Space Systems and Northrop Grumman Corporation. Special acknowledgements to Mr. Peter Klupar, David Mayer and Brett Pugh for their support on this proposal and to the Science Team.

## REFERENCES

[1] Mayor, M., and Queloz, D., "A Jupiter-mass companion to a solar-type star," *Nature* 378 (6555), 355–359 (1995).

[2] Han, E., et al., "Exoplanet Orbit Database. II. Updates to Exoplanets.org," PASP Vol 126, issue 943, pp 827-837 (2014).

[3] Stapelfeldt, K., et. al., "*Exo-C Interim Report,*" (2014).

[4] Seager, C. S., et. al., "*Probe Class Starshade Mission STDT Progress Report,*" (2014).

[5] Kopparapu, R. K., Ramirez, R., Kasting, J. F., Eymet, V., Robinson, T. D., Mahadevan, S., Deshpande, R., "Habitable zones around main-sequence stars: New Estimates," *ApJ, 765* (2013).

[6] Holman, M., & Wiegert, P. A., "Long-Term Stability of Planets in Binary Systems," *ApJ, 117*, 621-628 (1999).

[7] Guyon, O., "Phase-Induced Amplitude Apodization of Telescope Pupils for Extrasolar Terrestrial Planet Imaging," *Astronomy & Astrophysics*, 379-387 (2003).

[8] Belikov, R., Kasdin, N. J., & Vanderbei, R. J., "Diffraction-based Sensitivity Analysis of Apodized Pupil-mapping Systems. *The Astrophysical Journal,*" *652*, 833-844 (2006).

[9] Pluzhnik, E. A., Guyon, O., Ridgway, S. T., Martinache, F., Woodruff, R. A., Blain, C., & Galicher, R., "Exoplanet Imaging with a Phase-induced Amplitude Apodization Coronagraph. III. Diffraction Effects and Coronagraph Design," *The Astrophysical Journal, 644*, 1246-1257 (2006).

[10] Give'on, A., Belikov, R., Shaklan, S., & Kasdin, J., "Closed loop, DM diversity-based, wavefront correction algorithm for high contrast imaging systems," *Optics Express, 15*(19), 12338-12343 (2007).

[11] Bendek, E., Thomas, E., Belikov, R., "Direct Imaging of Exoplanets around Alpha Centauri and Other Multiple Star Systems," Proc. 014ebi Conf, 415B, (2014).

[12] Thomas, S., Bendek, E., Belikov, R., "Simulation of a method to directly image exoplanets around multiple stars systems," Proc. SPIE 9143 Optical Infrared and Millimeter Wave, 914335, (2014)

[13] Thomas, S., Belikov, R., Bendek, E., "Techniques for High Contrast Imaging in Multi-Star Systems I: Super-Nyquist Wavefront Control," *ApJ,* (2014).

[14] Lozi, J., Belikov, R., Thomas, S. J., Pluzhnik, E., Bendek, E., Guyon, O., & Schneider, G., "Experimental study of a low-order wavefront sensor for high-contrast coronagraphic imagers: results in air and in vacuum," *Proc. SPIE 9143* (2014).

[15] Wilkins, A., N., McElwain, M., W., Norton, T., J., Rauscher, B., J., Rothe, J., F., Malatesta, M., Hilton, G., M., Bubeck, J., R., Grady, C., A., Lindler, D., J., "Characterization of a photon counting EMCCD for space-based high contrast imaging spectroscopy of extrasolar planets," Proc. of the SPIE, Volume 9154, 91540C, 12 pp (2014).

[16] Jared R. Males, Ruslan Belikov, Eduardo A. Bendek, "Orbital difference imaging: a new high-contrast post-processing technique for direct imaging of exoplanets," Proc. SPIE 9605-42 (2015).

[17] Soummer, R., Pueyo, L., & Larkin, J., "Detection and Characterization of Exoplanets and Disks using Projections on Karhunen-Loeve Eigenimages," ApJ, 755:L28 (2012).